\documentstyle[twoside,11pt]{article}
\normalsize
\def\greaterthansquiggle{\raise.3ex\hbox{$>$\kern-.75em\lower1ex\hbox{$\sim$}}}
\def\lessthansquiggle{\raise.3ex\hbox{$<$\kern-.75em\lower1ex\hbox{$\sim$}}}
\newcommand{\bdi}{\begin{displaymath}}
\newcommand{\edi}{\end{displaymath}}
\newcommand{\bfi}{\begin{figure}}
\newcommand{\efi}{\end{figure}}

\newcommand{\beq}{\begin{equation}}
\newcommand{\eeq}{\end{equation}}

\newcommand{\beqa}{\begin{eqnarray}}
\newcommand{\eeqa}{\end{eqnarray}}
\newcommand{\no}{\nonumber}

\newcommand{\ra}{\rightarrow}

\def\au{{\setbox0=\hbox{\lower1.36775ex%
\hbox{''}\kern-.05em}\dp0=.36775ex\hskip0pt\box0}}
\def\ao{{}\kern-.10em\hbox{``}}

\newcommand{\ddsla}{\partial\hspace{-4.6pt} /  }

\newcommand{\AAsla}{A\hspace{-5pt}  /  }

\begin{document}
\bibliographystyle{plain}

\begin{titlepage}
\begin{flushright}
UWThPh-1996-16\\
\today
\end{flushright}
\vspace{2cm}
\begin{center}
{\Large \bf General bound-state structure of the massive Schwinger model}\\[1cm]
C. Adam* \\
Institut f\"ur Theoretische Physik \\
Universit\"at Wien 
\vfill
{\bf Abstract} \\
\end{center}

Within the Euclidean path integral and mass perturbation theory we derive,
from the Dyson-Schwinger equations of the massive Schwinger model, a
general formula that incorporates, for sufficiently small fermion mass, 
all the bound-state mass poles of the massive Schwinger model. As an
illustration we perturbatively compute the masses of the three lowest
bound states.

\vfill

$^*)${\footnotesize 
email address: adam@pap.univie.ac.at}
\end{titlepage}

\section{Introduction}

The massless Schwinger model - which is two-dimensional QED with one massless 
fermion -  is wellknown to be exactly soluble (\cite{Sc1} - \cite{SW1}, 
\cite{Adam} - \cite{ABH}), and its solution may be used as a starting point 
for a (fermion) mass perturbation theory of the massive Schwinger model
(\cite{MSSM}, \cite{SMASS}, \cite{Co1}, \cite{FS1}). In both models 
instantons and a nontrivial vacuum structure ($\theta$-vacuum) are present
(\cite{DSEQ} - \cite{Sm2}).  
The spectrum of the massless model consists of one {\em free}, massive boson 
with Schwinger mass $\mu_0^2 =\frac{e^2}{\pi}$ 
(fermion-antifermion bound state \cite{IP1}, \cite{CKS})
and trivial higher states consisting of $n$ free Schwinger bosons.

For the massive model these higher states turn into $n$-boson bound states.
Their masses, in principle, could be computed, using mass perturbation theory, 
by evaluating the mass poles of the corresponding $n$-point functions.
Here we will adapt a slightly different method. 
By exploiting the Dyson-Schwinger
equations of the model we will find that all bound-state mass poles are
contained within one formula. From this we will compute the masses of the three 
lowest bound states perturbatively.

All computations are performed within the Euclidean path integral formalism 
and are done for general vacuum angle $\theta$. This latter fact causes some
minor complications, because for $\theta\not= 0$ parity is no longer 
conserved and, as a consequence, the mass pole equations will turn into
matrix equations.

\section{Massless Schwinger model}

Before starting the actual computations, we need some formulae from the massless
Schwinger model. Indeed, the
vacuum functional and Green functions of the massive Schwinger model in mass 
perturbation theory may be traced back to space-time integrations of VEVs
of the massless model,
\beq
Z(m,\theta)=\sum_{k=-\infty}^{\infty}e^{ik\theta}Z_k (m)
\eeq
($k$ \ldots instanton number) where
\bdi
Z_k (m)=N\int D\bar\Psi D\Psi DA^\mu_k \sum_{n=0}^\infty \frac{m^n}{n!}
\prod_{i=1}^n
\int dx_i \bar\Psi (x_i)\Psi (x_i)\cdot
\edi
\beq
\cdot e^{\int dx\Bigl[ \bar\Psi(i\ddsla
-e\AAsla_k )\Psi -\frac{1}{4}F_{\mu\nu}F^{\mu\nu}\Bigr] }
\eeq
and the mass perturbation expansion is yet performed. A general VEV of the 
massive model is given by
\beq
\langle \hat O \rangle_m =\frac{1}{Z(m,\theta)} \langle \hat O
\sum_{n=0}^\infty \frac{m^n}{n!}\prod_{i=1}^n \int dx_i \bar\Psi (x_i) \Psi
(x_i) \rangle_0 .
\eeq
For these expressions we need (pseudo-) scalar and vectorial VEVs of the 
massless model.
It is useful to rewrite the scalar densities in terms of chiral
ones, $S(x)=S_+ (x)+S_- (x)$, $S_\pm \equiv \bar\Psi P_\pm \Psi$, because for
VEVs of chiral densities
only a definite instanton sector $k=n_+ -n_-$ contributes,

\newpage
 
\beq
\langle S_{H_1}(x_1)\cdots S_{H_n}(x_n)\rangle_0 = e^{ik\theta} 
\Bigl( \frac{\Sigma}{2}\Bigr)^n \exp
\Bigl[ \sum_{i<j}(-)^{\sigma_i \sigma_j}4\pi D_{\mu_0} (x_i -x_j)\Bigr] 
\eeq
(see e.g. \cite{DSEQ}--\cite{Diss}, \cite{Zah}, \cite{MSSM} for the
computation)
where $\sigma_i =\pm 1$ for $H_i =\pm$, $D_{\mu_0}$ is the massive scalar
propagator, 
\beq
D_{\mu_0}(x)=-\frac{1}{2\pi}K_0 (\mu_0 |x|), \quad \widetilde
D_{\mu_0}(p)= \frac{-1}{p^2 +\mu_0^2},
\eeq
($K_0\ldots$ McDonald function) and $\Sigma$ is the fermion condensate
of the massless Schwinger model,
\beq
\Sigma =\langle\bar\Psi \Psi\rangle_0 =\frac{e^\gamma}{2\pi}\mu_0
\eeq
($\gamma\ldots$ Euler constant). 
An inclusion of an arbitrary number of vector currents does not alter the
contributing instanton sector and may be computed from the generating
functional
\bdi
\langle S_{H_1}(x_1)\cdots S_{H_n}(x_n)\rangle_0 [\beta]= e^{ik\theta} 
\Bigl( \frac{\Sigma}{2}\Bigr)^n \exp
\Bigl[ \sum_{i<j}(-)^{\sigma_i \sigma_j}4\pi D_{\mu_0} (x_i -x_j)\Bigr] \cdot 
\edi
\beq
\cdot \exp\Bigl[\int dy_1 dy_2 \beta (y_1) D_{\mu_0}(y_1 -y_2)\beta (y_2)
+2\sqrt{\pi}\sum_{l=1}^n (-)^{\sigma_l} \int dy\beta (y)D_{\mu_0}(y-x_l)
\Bigr]. 
\eeq
More precisely, (7) generates all VEVs of $n$ chiral densities and
an arbitrary number of Schwinger bosons $\phi$, where $\phi$ is
related to the vector current via
\beq
J_\mu =\frac{1}{\sqrt{\pi}}\epsilon_{\mu\nu}\partial^\nu \phi .
\eeq
(7) may be found by the inclusion of a vector current source into the path
integral quantization and was explicitly computed in \cite{Gatt}.

\input psbox.tex

\section{The bound-state mass poles}

First we have to fix some notation for later convenience:
\bdi
E_\pm (x):= e^{\pm 4\pi D_{\mu_0}(x)} - 1
\edi
\bdi
E^{(n)}_\pm (x) := e^{\pm 4\pi D_{\mu_0}(x)} - \sum_{l=0}^n \frac{1}{l!}
(\pm 4\pi D_{\mu_0}(x))^l
\edi
\beq
\widetilde E^{(n)}_\pm (p) = \int d^2 xe^{ipx}E^{(n)}_\pm (x) \quad ,\quad
E^{(n)}_\pm := \widetilde E^{(n)}_\pm (0).
\eeq
We will use the following Feynman rules:

$$\psboxscaled{700}{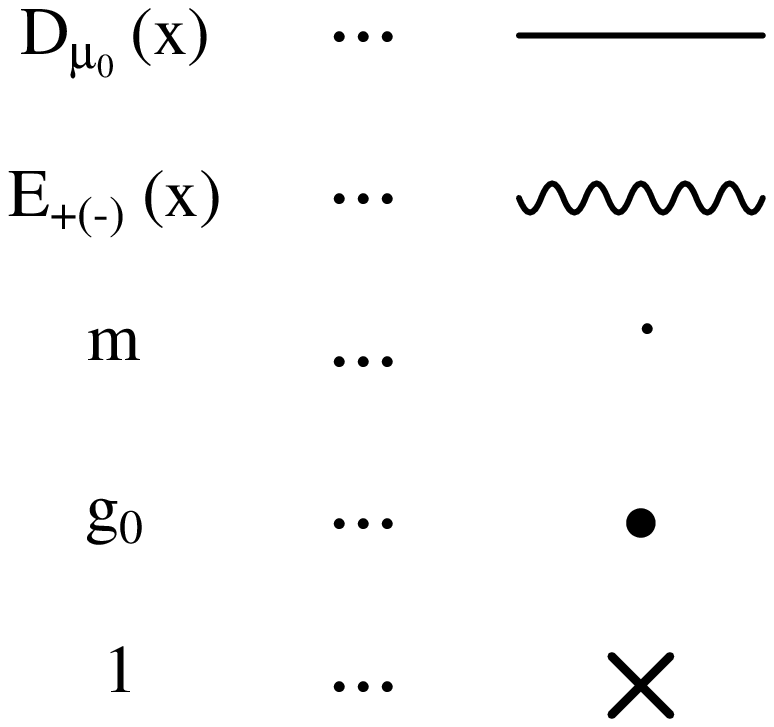}$$

\begin{center}
Fig. 1
\end{center}

\medskip

where $m$ and $g_0 = m\langle S(x)\rangle_m $ are the bare and renormalized 
coupling, respectively. In the sequel all VEVs are with respect to the
massive model, therefore we will omit the subscript $m$. 

We will discuss the special case $\theta =0$ first, because it is easier
and may be represented by simple graphical computations. Later we
generalize to arbitrary $\theta$.

On fermionic bilinears there hold two equations of motion, namely the Maxwell
equation and the anomaly equation. Eliminating the field strength one
arrives at
\beq
(\Box_x - \mu_0^2 )\phi (x) =2\sqrt{\pi}mP(x) \quad ,\quad P=S_+ -S_-
\eeq
where the Schwinger boson $\phi$ is related to the vector current like in
(8). Introducing the abbreviation
\beq
M_x := \Box_x -\mu_0^2
\eeq
one may derive Dyson-Schwinger equations like e.g. for the two-point function,
\bdi
M_{y_1}M_{y_2}\langle\phi (y_1)\phi (y_2)\rangle = M_{y_1}\delta (y_1 -y_2)+
\edi
\beq
4\pi g_0 \delta (y_1 -y_2) +4\pi g_0^2 \langle P(y_1)P(y_2)\rangle ,
\eeq
where it is understood that external sources couple with coupling constant 1.
The validity of equation (12) is most easily seen in a graphical representation
for the two-point function,

$$\psboxscaled{800}{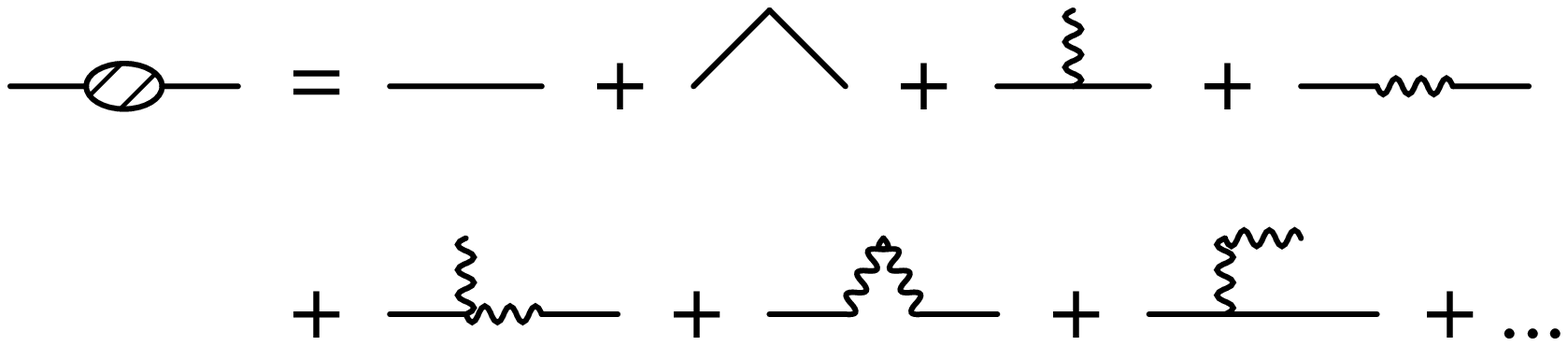}$$

\begin{center}
Fig. 2
\end{center}

\medskip

Here all graphs where the two boson lines meet on one point contribute to 
the renormalized coupling,

$$\psboxscaled{1000}{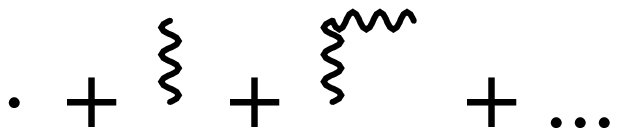}$$

\begin{center}
Fig. 3
\end{center}

\medskip

\beq
g_0 \equiv m\langle S\rangle =m\Sigma + \frac{1}{2}(m\Sigma)^2
(E_+ +E_-) +\ldots ,
\eeq
see \cite{MSSM}, \cite{SMASS} for computational details. All internal vertices
are renormalized by the same coupling $g_0$, and the remaining 
$\langle P(y_1)P(y_2)\rangle $ part may be written like 

$$\psboxscaled{900}{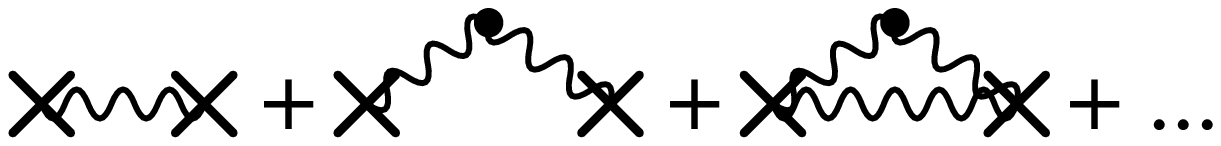}$$

\begin{center}
Fig. 4
\end{center}

\medskip

The $\langle P(y_1)P(y_2)\rangle $ propagator includes, even in least order, 
an arbitrary number of bosons propagating from $y_1$ to $y_2$, which will
be essential in the following.

Similar Dyson-Schwinger equations may be derived for higher $n$-point functions,
e.g. for the four-point function
\bdi
M_{y_1} M_{y_2} M_{y_3} M_{y_4} \langle \phi (y_1)\phi (y_2)\phi (y_3)
\phi (y_4)\rangle =
\edi
\bdi
M_{y_1} M_{y_2} \langle \phi (y_1)\phi (y_2)\rangle
M_{y_3} M_{y_4} \langle \phi (y_3)\phi (y_4)\rangle 
+\mbox{ perm. }+
\edi
\bdi
16\pi^2 g_0 \delta (y_1 -y_2)\delta (y_1 -y_3)\delta (y_1 -y_4) +
\edi
\bdi
16\pi^2 g_0^2 \delta (y_1 -y_2) \delta (y_3 -y_4) \langle S(y_1)S(y_3)
\rangle + \mbox{ perm. } +
\edi
\bdi
16\pi^2 g_0^3 \delta (y_1 -y_2)\langle S(y_1)P(y_3)P(y_4)\rangle +
\mbox{ perm. } +
\edi
\beq
16\pi^2 g_0^4 \langle P(y_1)P(y_2)P(y_3)P(y_4)\rangle
\eeq
and analogously for higher $n$-point functions.

The essential point is that in all these Dyson-Schwinger equations there
occurs an identical term that will be responsible for the bound-state formation
for sufficiently small fermion mass. In momentum space this term reads
\beq
c(g_0 +g_0^2 \widetilde{\langle PP\rangle}(p))
\eeq
for odd bound states ($c=4\pi ,16\pi^2 ,\ldots$) and
\beq
c(g_0 +g_0^2 \widetilde{\langle SS\rangle}(p))
\eeq
for even bound states.

Now both terms (15), (16) may be inverted via the geometric series 
formula, e.g.
\beq
g_0 (1+g_0 \widetilde{\langle PP\rangle}(p))=\frac{g_0}{1-g_0
\widetilde{\langle PP\rangle}_{\rm n.f.}(p)}
\eeq
where n.f. stands for nonfactorizable and means that graphs contributing to
$\widetilde{\langle PP\rangle}_{\rm n.f.}$ may not be factorized in momentum
space. Its graphical representation looks like

$$\psboxscaled{900}{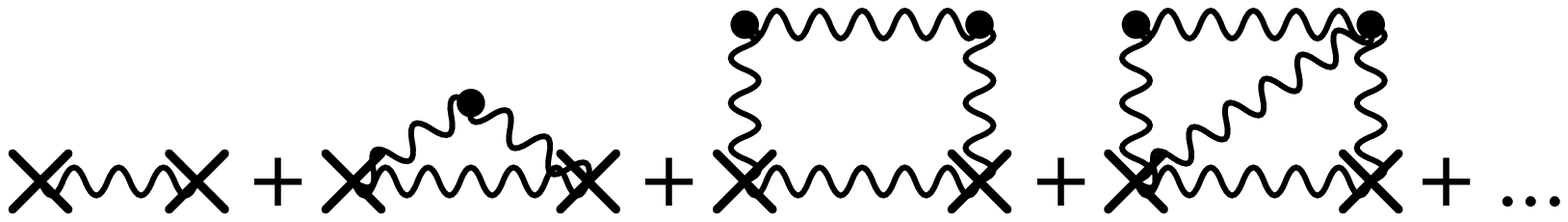}$$

\begin{center}
Fig. 5
\end{center}

\medskip

In (17) the small $g_0$ has to be compensated by a large contribution from
$\widetilde{\langle PP\rangle}_{\rm n.f.}$ in order to give rise to a mass pole.
The first term in Fig. 5 is $E_\pm (x)=e^{\pm 4\pi D_{\mu_0}(x)}-1$
and contains an arbitrary number of massive propagators $D_{\mu_0}$.
Now precisely $(D_{\mu_0}(x))^n$ has a threshold singularity in momentum
space at $p^2 =(n\mu_0 )^2$, therefore (17) may have mass poles near $p^2=
(n\mu_0 )^2$ for $n\in {\rm\bf N}$. More precisely, rewriting (for $\theta =0$)
$\langle PP\rangle =2\langle S_+ S_+ \rangle - 2\langle S_+ S_- \rangle $
(and with a + for $\langle SS\rangle $), and using
\beq
\langle S_+ (x)S_+ (0)\rangle =\Bigl(\frac{\Sigma}{2}\Bigr)^2
e^{+4\pi D_{\mu_0}(x)} \quad ,\quad
\langle S_+ (x) S_- (0)\rangle = \Bigl( \frac{\Sigma}{2}\Bigr)^2
e^{-4\pi D_{\mu_0}(x)}
\eeq
we find that $\widetilde{\langle PP\rangle}$ may cause mass poles for {\em odd} $n$
whereas $\widetilde{\langle SS\rangle}$ may cause mass poles for {\em even} $n$,
as it has to be. For all mass poles only the terms (15), (16) may
balance the pole equation for sufficiently small $g_0$, therefore it is
enough to consider them.

To get more insight we next have to rewrite
$\widetilde{\langle PP\rangle}_{\rm n.f.}$ (Fig. 5) in terms of internal
Schwinger bosons (we ignore constants)
 
$$\psboxscaled{900}{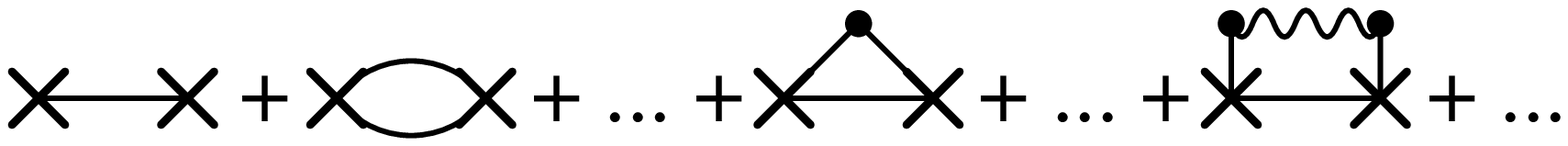}$$

\begin{center}
Fig. 6
\end{center}

\medskip

We find that the one-boson propagator acquires {\em no} corrections, whereas 
all terms with more than one boson propagating from $y_1$ to $y_2$ have
corrections along the boson lines. Consequently, one can compute the lowest
pole mass (the Schwinger boson mass) without the need to know it.

On the other hand, for the computation of higher bound-state masses,
one has to take into account the corrections, i.e. insert the exact Schwinger
mass (the lowest pole mass). The reason is that the mass corrections
for the bosons just shift the position of the threshold singularity and
are therefore important in lowest order. There are other corrections present,
too (internal boson interactions), however, they are unimportant in lowest
order. 

This result is very plausible physically: the higher bound states should
consist of {\em physical} Schwinger bosons with their physical masses $\mu$
(not the bare masses $\mu_0$).

All these features remain true for general $\theta$, only the pole mass
formula itself is slightly more complicated and shall be derived next.

\section{General $\theta$ case}

For $\theta \ne 0$ the renormalized coupling is complex,
\beq
g=m\langle S_+ \rangle \quad ,\quad g^* =m\langle S_- \rangle
\eeq
and the Dyson-Schwinger equations are slightly changed, too. E.g. the 
two-point function obeys
\bdi
M_{y_1} M_{y_2} \langle \phi (y_1)\phi (y_2)\rangle = M_{y_1}
\delta (y_1 -y_2) + (g+g^*)\delta (y_1 -y_2) +
\edi
\beq
g^2 \langle S_+ (y_1)S_+ (y_2)\rangle + (g^* )^2 \langle S_- (y_1)S_- (y_2)
\rangle -2gg^* \langle S_+ (y_1) S_- (y_2)\rangle .
\eeq
The interesting function that gives rise to the mass poles is
\beq
g(1+g\widetilde{\langle S_+ S_+ \rangle}(p) - g^* \widetilde{\langle S_+ S_- \rangle}
(p)) + g^* (1 + g^* \widetilde{\langle S_- S_- \rangle}(p) -g\widetilde{\langle
S_+ S_- \rangle}(p))
\eeq
for odd bound states and with only plus signs for even bound states.
Because parity is no longer conserved the $P$ and $S$ components mix and
the geometric series formula (17) generalizes to a matrix equation.

Introducing the abbreviations
\bdi
A:=1+g\widetilde{\langle S_+ S_+ \rangle}(p) - g^* \widetilde{\langle S_+ S_- \rangle}
(p)
\edi
\beq
\alpha :=g\widetilde{\langle S_+ S_+ \rangle}_{\rm n.f.}(p) \quad ,
\quad \beta :=g\widetilde{\langle S_+ S_- \rangle}_{\rm n.f.}(p)
\eeq
the equation reads
\beqa
A&=&1+\alpha A-\beta^* A^* \no \\
A^* &=& 1+\alpha^* A^* -\beta A
\eeqa
and has the solution
\beq
A=\frac{1-\alpha^* -\beta^* }{1-\alpha -\alpha^* +\alpha\alpha^* 
-\beta\beta^* }.
\eeq
Equation (23) may be checked by a careful investigation of the perturbative
expansion for $\langle P(y_1)P(y_2)\rangle$.

For even bound states the solution may be found from (24) by the
substitution $\beta \ra -\beta$. Therefore, all solutions have the same 
denominator, and the zeros of this denominator are the pole masses of all 
bound states of the massive Schwinger model.

Explicitly the pole-mass equation reads
\beq
(1-g\widetilde{\langle S_+ S_+ \rangle}_{\rm n.f.}(p))(1-g^* \widetilde{\langle S_+
S_+ \rangle}^*_{\rm n.f.} (p))=gg^* (\widetilde{\langle S_+ S_- \rangle}_{\rm
n.f.}(p))^2
\eeq
where
\beqa
g &=& m\frac{\Sigma}{2}e^{i\theta} + m^2 \Bigl(\frac{\Sigma}{2}\Bigr)^2
(E_+ e^{2i\theta} +E_- )+o(m^3) \no \\
&=:& g_1 +g_2 +o(m^3)
\eeqa
\beq
\widetilde{\langle S_+ S_+ \rangle}_{\rm n.f.}(p) =\widetilde E_+ (p)+o(m)\quad ,
\quad \widetilde{\langle S_+ S_- \rangle}_{\rm n.f.}(p) =\widetilde E_- (p)+o(m) .
\eeq
Of course, for $\theta =0$ (real $A,\alpha ,\beta$), one recovers the
equations (15), (16).

\section{Explicit mass computations}

For a computation of the Schwinger mass up to second order we rewrite (25)
like
\beq
(1-g\widetilde E_+ (p))(1-g^* \widetilde E_+ (p))=gg^* \widetilde E_-^2 (p)
\eeq
and separate the one-boson contribution
\beq
\widetilde E_\pm (p)=-\frac{\pm 4\pi}{p^2 +\mu_0^2}+\widetilde E_\pm^{(1)} (p)
\eeq
leading to
\beq
-p^2 -\mu^2_0 =8\pi {\rm Re}\, g_1 -2{\rm Re}\, g_1 \widetilde E_+^{(1)} (p)
(p^2 +\mu_0^2 )+8\pi {\rm Re}\, g_2 +8\pi g_1 g_1^* (\widetilde E_+^{(1)} (p)
-\widetilde E_-^{(1)} (p))
\eeq
with the solution
\beq
-p^2 =\mu_0^2 (1+4\pi \frac{\Sigma m}{\mu_0^2}\cos\theta )
\eeq
in first order and
\beq
-p^2 =\mu_0^2 \Bigl[ 1+4\pi \frac{\Sigma m}{\mu_0^2}\cos\theta +2\pi \frac{
m^2 \Sigma^2}{\mu_0^4}\Bigl( ( E_+ + \widetilde E_+^{(1)} (1))\cos 2\theta +
E_- -\widetilde E_+^{(1)} (1)\Bigr) \Bigr]
\eeq
in second order. Here we rescaled $p\ra p' =\frac{p}{\mu_0}$ and used
$\widetilde E_\pm^{(1)} (p' )=\widetilde E_\pm^{(1)} (1)+o(m)$ in the last step.
This result precisely conicides with the result obtained by a direct 
perturbative computation (\cite{SMASS}).

In order to compute the two-boson bound state we have to separate the 
two-boson part of $\widetilde E_+$ in (28). In lowest order we find
\beq
1=\frac{1}{2!}(g_1 +g_1^* )16\pi^2 \widetilde{(D_\mu^2 )} (p)
\eeq
where now $\mu$ is the {\em physical} Schwinger mass (32) including 
fermion mass corrections. Using
\beq
\widetilde{(D_\mu^2 )} (p) =\frac{1}{4\pi 
(-p^2)}\frac{1}{\sqrt{\frac{4\mu^2}{-p^2}-1}}
\arctan \frac{1}{\sqrt{\frac{4\mu^2}{-p^2}-1}}
\eeq
(see e.g. \cite{BOUND} for a computation) and remembering that $4\mu^2 -(-p^2)$
is a very small number (that is {\em positive} for a bound state) we may set
$\frac{1}{-p^2}\simeq \frac{1}{4\mu^2}$, $\arctan (\cdots )\simeq\frac{\pi}{2}$ 
and get
\beq
-p^2 \simeq 4\mu^2 (1-\frac{\pi^4}{16}\frac{m^2 \Sigma^2}{\mu^4}\cos^2 \theta )
\eeq
which is of second order in $m$. Again, this result coincides with the one
from a direct perturbative calculation (\cite{BOUND}).

\section{The three-boson bound-state mass}

For the three-boson bound-state mass we have to separate the three-boson part
in (28) and find, in lowest order
\beq
1=\frac{1}{3!}m\Sigma \cos \theta \cdot 64\pi^3 \widetilde{(D_\mu^3 )} (p)
\eeq
or, after a rescaling $p\ra \frac{p}{\mu}$ to dimensionless momenta
\beq
1=\frac{64\pi^3}{6}\frac{m\Sigma}{\mu^2}\cos\theta
\, \widetilde{(D_\mu^3 )} (p).
\eeq
$\widetilde{(D_\mu^3 )} (p)$ is given by the graph (where we introduce positive 
squared momentum $Q^2 =-p^2 > 0$)

$$\psboxscaled{1000}{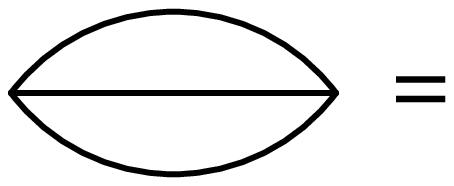}$$

\begin{center}
Fig. 7
\end{center}

\medskip

\bdi
-\int \frac{d^2 q_1 d^2 q_2}{(2\pi)^4}\frac{1}{(p+q_1 +q_2)^2 +1}
\frac{1}{q_1^2 +1}\frac{1}{q_2^2 +1}=
\edi
\bdi
-2\int_0^1 dx\int_0^x dy\int  \frac{d^2 q_1 d^2 q_2}{(2\pi)^4}
\frac{1}{\Big[ q_1^2 +1+(q_2^2 -q_1^2)x + ((p+q_1 +q_2 )^2 -q_2^2)y\Bigr]^3}=
\edi
\bdi
\int \frac{dx}{(4\pi)^2}\int_0^x \frac{dy}{Q^2 (xy-x^2 y-y^2 +xy^2 )-x+x^2 -xy 
+y^2} =
\edi
\bdi
\int \frac{dx}{8\pi^2 (1-Q^2 (1-x))}\int_0^{\frac{x}{2}}
\frac{dz}{z^2 +T^2 (Q^2 ,x)} =
\edi
\beq
\int_0^1 \frac{dx}{8\pi^2 (1-Q^2 (1-x))}\frac{1}{T(Q^2 ,x)}
\arctan \frac{x}{2T(Q^2 ,x)} ,
\eeq
where
\beq
T^2 (Q^2 ,x)=\frac{x^2 -Q^2 x^2 (1-x)+4x(1-x)}{4(Q^2 (1-x)-1)} .
\eeq
The numerator of $T^2$ has a double zero at $Q^2 =9$:
\beq
9x(x-\frac{2}{3})^2 .
\eeq
This double zero is in the integration range of $x$ and is precisely the 
threshold singularity. Setting
\beq
Q^2 =:9(1-\epsilon )
\eeq
in the numerator of $T^2$ in the factor $\frac{1}{T}$, and $Q^2 =9$ everywhere
else, where it is safe, one arrives at:
\beq
\frac{1}{12\pi^2}\int_0^1 \frac{dx}{\sqrt{\vert 9x-8\vert }}
\frac{\arctan \frac{\sqrt{x\vert 9x-8\vert }}{3(x-\frac{2}{3})}}{
\sqrt{(x-\frac{2}{3})^2 x+\epsilon x^2 (1-x)}}=:I(\epsilon ).
\eeq
The mass-pole equation reads ($\frac{m\Sigma}{\mu^2}\cos\theta \equiv
\frac{1}{\mu^2}2{\rm Re}\, g_1 =:\alpha$)
\beq
1=\frac{64\pi^3}{6}\alpha I(\epsilon )
\eeq
and must be evaluated numerically. It gives rise to an extremely tiny mass 
correction $\epsilon$. For sufficiently small $\alpha$ it is very well 
saturated by
\beq
\epsilon (\alpha )\simeq 0.777 \exp (-\frac{0.263}{\alpha})
\eeq
and is therefore smaller than polynomial in $\alpha$. (I checked the 
numerical formula (44) for $30 <\frac{1}{\alpha}<1000$, corresponding to
$10^{-3}<\epsilon <10^{-100}$, but I am convinced that it remains true for
even larger $\frac{1}{\alpha}$; however, there the numerical integration
is quite difficult because of the pole in (42).) 

A more accurate mass-pole equation would include some additional contributions:
\beq
1=\alpha (\frac{64\pi^3}{6}I(\epsilon )+f(\epsilon ))
\eeq
where $f(\epsilon )$ is some function that is finite for $\epsilon \to 0$.
But for $\alpha$ sufficiently small it remains true that an extremely tiny 
value of $\epsilon$ suffices to saturate the mass-pole equation,
whatever the value of $f(\epsilon )$ is.

We conclude that the three-boson bound state mass is nearly entirely given by 
three times the Schwinger boson mass,
\beq
M_3^2 \simeq 9\mu^2 ,
\eeq
or, differently stated, that the binding of three bosons is extremely weak.

This result enables us to add a short remark on a result that was obtained
in \cite{Co1}. There it was argued that, for $\theta =0$, the three-boson
bound state should be stable because a decay into two Schwinger bosons
is forbidden by parity conservation. However, because of (46) it holds that
$M_3 >M_2 +\mu$, therefore there should be a small probability for the
three-boson bound state to decay into one two-boson bound state and one
Schwinger boson.

\section{Summary}

As claimed, we arrived at our aim to derive one formula for all bound-state 
masses of the massive Schwinger model, at least for sufficiently small
fermion mass. Of course, if we were able to exactly solve the model, it
would not at all be surprizing to find all mass poles within one Green
function. The interesting point is that we could reach this aim by
the use of the Dyson-Schwinger equations and by a specific partial
resummation of the perturbation series.

We found that the two lowest states acquire noticeable corrections, whereas
the binding energy of the three-boson bound state is extremely tiny.
It is plausible to conjecture that the binding energies of higher bound
states remain very tiny.

\section*{Acknowledgement}

The author thanks M. Hutter for intensive and very helpful discussions.

In addition, the author thanks Prof. Narison for 
the invitation to the Institute of 
Theoretical Physics of the University of Montpellier, where part of this 
work was performed,
and the members of the institute for their hospitality. 

Further thanks are due to the French Foreign Ministery, the Austrian
Ministery of Research and the Austrian Service for Foreign Exchange, who
financially supported this research stay.

\end{document}